\documentclass{procDDs}
\usepackage{color}

\title{Asymptotics of Green function for the linear waves equations in a domain with a non-uniform bottom}

\author{A.ANIKIN, S.DOBROKHOTOV, V.NAZAIKINSKII}
{Ishlinski Institute for Problems of Mechanics and 
Moscow Institute of Physics and Technology, Moscow, Russia}                 
{anikin83@inbox.ru~; dobr@ipmnet.ru~; nazay@ipmnet.ru}                                   
\coauthor{M.ROULEUX}                       
{Aix Marseille Univ, Univ Toulon, CNRS, CPT, Marseille, France}               
{rouleux@univ-tln.fr}                                               

\def\e{\mathop{\rm \varepsilon}\nolimits}
\def\id{\mathop{\rm Id}\nolimits}

\def\Op{\mathop{\rm Op}\nolimits}
\def\re{\mathop{\rm Re}\nolimits}

\begin {document}
\newtheorem{theo}{Theorem}
\newtheorem{prop}{Proposition}
\newtheorem{definition}{Definition}
\maketitle

\index{Author1, I.I.}                              
\index{Author2, I.I.}                              
\index{Coauthor, I.I.}                             %

\begin{abstract}
We consider the linear problem for water-waves 
created by sources on the bottom and the free surface 
in a 3-D basin having slowly varying profile $z=-D(x)$. The fluid
verifies Euler-Poisson equations.
These (non-linear) equations have been given a Hamiltonian form by Zakharov,
involving canonical variables $(\xi(x,t),\eta(x,t))$ 
describing the dynamics of the free surface; variables $(\xi,\eta)$ are related by the free surface
Dirichlet-to-Neumann (DtN) operator. For a single variable $x\in{\bf R}$ and constant depth, DtN operator was explicitely
computed in terms of a convergent series. Here we neglect quadratic terms in Zakharov equations,
and consider the linear response to a disturbance of $D(x)$ harmonic in time when the wave-lenght is small compared to
the depth of the basin. We solve the Green function problem for a matrix-valued DtN operator, at the bottom and the free-surface.  
\end{abstract}

\section{Euler-Poisson and Zakharov equations}

Consider the free surface of a 3-D fluid 
incompressible and irrotational, subject to the gravity field \cite{KuzMazVai}. The bottom profile is given by $z=-D(x,t)$, $x\in{\bf R}^2$.
The velocity potential $\Phi(x,z,t)$ satisfies $\Delta_{x,z}\Phi=0$
in the fluid region bounded by $z=-D(x,t)$ and the free surface $z=\eta(x,t)$ (thus delimiting a domain $\Omega(t)\subset{\bf R}^3$), 
with the boundary condition
$\partial_{\vec n}\Phi=v_{\vec n}$ at $z=-D(x,t)$ 
where $\vec n=\vec n(x,t)$ denotes the outer normal to $\Omega(t)$ at the given point $x$, 
and $v_{\vec n}$ the time-dependent normal velocity of the bottom.
On the free surface we assume the kinematic boundary condition and Bernoulli law
\begin{equation}
\label{0.2}
\begin{aligned}
&\partial_t\eta+\langle\nabla_x\Phi,\nabla_x\eta\rangle-\partial_z\Phi=0 \ @ \ z=\eta(x,t)\cr
&\partial_t\Phi+{1\over2}|\nabla\Phi|^2+g\eta+{p\over\rho}=0 \ @ \ z=\eta(x,t)\cr
\end{aligned}
\end{equation}
respectively, where $p=p(x,t)$ stands for pressure, which we consider here as a source term. 
Let $\xi(t,x)=\Phi(x,\eta(x,t),t)$ be the potential on the free surface. We define the free surface DtN map 
$G(\eta)$ at $z=\eta(x,t)$ so that \ref{0.2} can be rewritten in Hamiltonian form (valued in infinite dimensional fields)
as follows \cite{Z}:
\begin{equation*}
\begin{aligned}
&\partial_t\eta-G(\eta)\xi=0\cr
&\partial_t\xi+g\eta+{p\over\rho}+{1\over2}|\nabla_x\xi|^2=\cr
&{1\over2}(1+|\partial_x\eta|^2)^{-1}
\bigl(\langle\nabla_x\xi,\nabla_x\eta\rangle+G(\eta)\xi\bigr)^2\cr
\end{aligned}
\end{equation*}
In 1-D case ($x\in{\bf R}$), and constant depth $D(x,t)=D$, 
DtN map $G(\eta)$ is computed as an operator valued, convergent Taylor expansion, of the form \cite{CrSu}
\begin{equation*}
\begin{aligned}
&G(\eta)=\sum_{j\geq0}G_j(\eta,DD_x)\cr
&G(0)\xi(x)=D_x\tanh(DD_x)\xi(x)
\end{aligned}
\end{equation*} 
with $D_x=-i\partial_x$. Our main goal is to extend semi-classically this formula to the 2-D linearized equations, for a variable depth profile,
and derive similarly the DtN operator at the bottom, and the DtN operator accounting for interaction
between the free surface and the bottom.
The linearized
set of scaled Euler-Poisson equations (with respect to the small parameter $h$) together with the equation for the potential, takes the form:
\begin{equation}
\label{0.13}
\begin{aligned}
&-h^2\Delta_x\Phi-{\partial^2\Phi\over\partial z^2}=0, \ -D(x)<z<0\cr
&h^2{\partial^2\Phi\over\partial t^2}+g{\partial\Phi\over\partial z}=f^+(x,t) \ @ \ z=0\cr
&{\partial \Phi\over\partial z}+h^2\langle\nabla_x D,\nabla_x\Phi\rangle=f^-(x,t) \ @ \ z=-D(x)\cr
\end{aligned}
\end{equation}
with given sources $f^\pm(x,t)$. Note that the ``moving'' boundaries $z=\eta(x,t)$ (resp. $z=-D(x,t)$) have also been frozen 
to $z=0$ (resp. $z=-D(x)$). 
Initial conditions (Cauchy data) must be further prescribed of the form:
$$\Phi|_{z=0,t=0}=\phi(x,t)|_{t=0}, \quad h{\partial\Phi\over\partial t}|_{z=0,t=0}=-\eta|_{t=0}$$
Here $\phi(x,t)=\Phi(x,z,t)|_{z=0}$ stands for (linearization of) $\xi(x,t)$ in Zakharov equations. 

For simplicity we look for stationary solutions with a time 
harmonic disturbance of the form $f^\pm(x,t)=e^{i\omega t/h}f^\pm_0(x)$ and the potential at the free surface 
of the form $\phi(x,t)=e^{i\omega t/h}\phi_0(x)$. 
We assume that the linear response is also harmonic with respect to time, with same frequency $\omega$.
Eq. \ref{0.13} then transform into a set of time-independent equations. See \cite{DoNaSh}, \cite{DoNa}, \cite{DoZh}, \cite{DoKuzZh}.

\section{Dirichlet-to-Neumann operator}

Consider here the problem of Laplace operator with Dirichlet condition at the free surface, and Neuman condition at the bottom
\begin{equation}
\label{1.4}
\begin{aligned}
&-h^2\Delta_{x}\Phi-{\partial^2\Phi\over\partial z^2}=f, \quad -D(x)<z<0\cr
&\Phi=\varphi^+(x) \ @ \ z=0\cr
&\partial_{\vec n_h}\Phi=\psi^-(x) \ @ \ z=-D(x)\cr
\end{aligned}
\end{equation}
together with the ``twin problem'' with Neumann condition at the free surface, and Dirichlet condition at the bottom.
\begin{equation}
\label{1.5}
\begin{aligned}
&-h^2\Delta_{x}\Psi-{\partial^2\Psi\over\partial z^2}=f, \quad -D(x)<z<0\cr
&{\partial\Psi\over\partial z}=\psi^+(x) \ @ \ z=0\cr
&\Psi=\varphi^-(x) \ @ \ z=-D(x)\cr
\end{aligned}
\end{equation}
Here $\vec n_h=\vec n_h(x)$ is the (scaled) outer normal of $\Omega=\{(x,z)\in{\bf R}^3: -D(x)<z<0\}$ at the given point. 
These equations may as well describe the heat diffusion in a fluid with $f$ as a source term (Poisson problem); in fact
Poisson problem would lead to consider DtN operator $G(\xi)$ at a threshold.
Denote as usual by
$\gamma^\pm_j$, $j=0,1$ the two first
trace operators on $\Gamma^+=\{z=0\}$, resp.
$\Gamma^-=\{z=-D(x)\}$. 
To avoid difficulties due to the fact that $x$ ranges over the infinite domain ${\bf R}^2$, 
we assume
\begin{equation}
\label{1.6}
\lim_{|x|\to\infty} D(x)=D_0>0 \ \hbox{in} \ C^2 \ \hbox{topology}
\end{equation}
Stereographic projection $T:{\bf R}^2\to{\bf S}^2$ then maps
problem \ref{1.4} onto a problem on the cylinder $X=\{(x,z):x\in {\bf S}^2, z\in[-D(x),0]\}$ with Neumann boundary condition on
$z=-D(x)$ and Dirichlet boundary condition on 
$z=0$. And similarly for \ref{1.5}. Recall \cite{Mel} the space of supported distributions 
$\dot{\cal D}'(X)$ dual to the space $C^\infty(X)$ of smooth functions up to $\partial X$,
and the space of extendible distributions  
${\cal D}'(X)$, dual to the space $\dot C^\infty(X)$ 
of smooth functions on $X$ vanishing of infinite order at $\partial X$.
The corresponding Sobolev spaces are denoted resp. by $\dot H^s(X)$ and $H^s(X)$. For all $s$ we have 
$\bigl[H^{-s}(X)\bigr]'\approx\dot H^s(X)$, and these spaces are reflexive. 
So let ${\cal H}\approx {\cal H}'=L^2(X)$,
$E=\dot H^1(X)$ and its dual space $E'=H^{-1}(X)$. We introduce the closed subspace
$F=\{v\in \dot H^1(X): \gamma_0^+(v)=0\}$ in $E$, and for $\varphi^+\in H^{1/2}(\Gamma^+)$ the closed affine subspace
$V_{\varphi^+}=\{v\in \dot H^1(X): \gamma_0^+(v)=\varphi^+\}$ with direction $F$.
Given $(f,\psi^-)\in H^{-1}(X)\times H^{-1/2}(\Gamma^-)$ 
Lax-Milgram (or Riesz-Fr\'echet) theorem
shows at once that there is a unique $u\in V_{\varphi^+}$ such that for all $v\in V_{\varphi^+}$
$$\int_{X}\nabla u\nabla v=\int_{X} fv+\int_{\Gamma^-}\psi^- v$$
which is the variational formulation of \ref{1.4}. It is then standard that for all
$(f,\psi^-,\varphi^+)\in L^2(\Omega)\times H^{1/2}(\Gamma^-)\times H^{3/2}(\Gamma^+)$,
\ref{1.4} has a unique solution $\Phi\in \dot H^2(\overline\Omega)$. We argue similarly 
for \ref{1.5}.
Next we define the generalized DtN map. For simplicity we assume $f=0$,
and consider
$${\bf L}{\varphi^+\choose\psi^-}={\psi^+\choose\varphi^-}$$
with ${\bf L}=\begin{pmatrix}L_{11}&L_{12}\\ L_{21}&L_{22} \end{pmatrix}$.
Physically operators $L=L_{11}$ and $L_{12}$ are most significant, since they give the linear response at the free surface, due to the 
disturbances both at the free surface and at the bottom. 
Moreover $L_{11}$ is the linearization of DtN map $G(\xi)$. 

It is readily seen (using Green formula) that operators 
$L_{11}:L^2(\Gamma^+)\to L^2(\Gamma^+)$ with domain  $H^{3/2}(\Gamma^+)$, and 
$L_{22}:L^2(\Gamma^-)\to L^2(\Gamma^-)$ with domain $H^{1/2}(\Gamma^+)$ are self-adjoint;
moreover $L_{21}^*=-L_{12}$.

\section{$h$-Pseudo-differential DtN operators}

We want to express operators $L_{ij}$ as $h$-PDO's on $L^2({\bf R}^2)$. 
Let $R(x,p;,h)$ be a smooth real classical Hamiltonian on $T^*{\bf R}^2$~; we will assume that
$R$ belongs to the space of symbols $S^0(m)$ for some order function $m$ (for example $m(x,p)= (1+|p|^2)^{1/2}$) with
\begin{equation}
\label{2.1}
\begin{aligned}
&S^N(m)=\{R\in C^\infty(T^*{\bf R}^d): \exists M,\forall\alpha,\beta\in{\bf N}^{2}\cr 
&\exists C_{\alpha\beta}>0, \ |\partial^\alpha_{x}\partial^\beta_pH(x,p;h)|\leq C_\alpha h^N m^M\}\cr
\end{aligned}
\end{equation}
We need include the case where the symbol $R$ depends also parametrically on $z$ 
(as a solution of a linear ODE) so we will to replace the estimate \ref{2.1} by
\begin{equation*}
\begin{aligned}
\forall&\alpha,\beta\in{\bf N}^{2}, \forall \ell\in {\bf N}, \exists C_{\alpha\beta\ell}>0, \exists M(\ell)\cr
&|\partial^\ell_z\partial^\alpha_x\partial^\beta_pH(x,p;z,h)|\leq C_{\alpha\beta\ell} h^N m^{M(\ell)}\cr
\end{aligned}
\end{equation*}
We shall mainly consider such symbols with the semi-classical expansion
$$H(x,p;z,h) \sim H_0(x,p;z)+hH_1(x,p;z)+\cdots, h\rightarrow 0$$

Recall first (Leibnitz formula for $h$-PDO's) that
$R(x,p+hD_x;h)\cdot 1=R(x,p;h)$ and more generally 
the composition of symbols $R\sharp S$ (for $qp$, or Feynman rule) is the symbol of 
$R(x,p+hD_x;h)\cdot S(x,p)$. We have
similar formulas when $R$ is valued in the class of differential operators with respect to the $z$-variable.
We seek the solution of \ref{1.4} in the form $\Phi=\widehat Ru$, where $\widehat R$ is a
$h$-PDO of the form above to be determined with symbol $R(x,p;z,h)$. 
It satisfies
\begin{equation}
\label{2.2}
\begin{aligned}
&\frac{\partial^2R}{\partial z^2}-(p+hD_x)^2R=0, \ -D(x)\leq z\leq0\cr \ 
&R|_{z=0}=1\cr
&\partial_zR+ih\langle\nabla D(x),(p+hD_x)R\rangle|_{z=-H(x)}=0\cr
\end{aligned}
\end{equation}
We look for $R$ in the form
$R(z,x,p,h)=R_0(z,x,p)-ih R_1(z,x,p)+\cdots$
and obtain by substitution
\begin{equation}
\label{2.3}
\begin{aligned}
&\frac{\partial^2R_0}{\partial z^2}-p^2R_0=0, \ -D(x)\leq z\leq0, R_0|_{z=0}=1\cr
&\frac{\partial R_0}{\partial z}|_{z=-D(x)}=0\cr
\end{aligned}
\end{equation}
and
\begin{equation}
\label{2.4}
\begin{aligned}
&\frac{\partial^2R_1}{\partial z^2}-p^2R_1=2\langle p,\frac{\partial R_0}{\partial x}\rangle, \ -D(x)\leq z\leq0\cr
&R_1|_{z=0}=0,\quad \frac{\partial R_1}{\partial z}|_{z=-D(x)}=\langle\nabla D,p\rangle B_0|_{z=-D(x)}\cr
\end{aligned}
\end{equation}
The solution of problem \ref{2.3} has the form
\begin{equation*}
R_0(x,p,z)=\frac{\cosh(z+H(x))|p|}{\cosh D(x)|p|}
\end{equation*}
Solving then analogously for $R_1$ in \ref{2.4} we end up with:

\begin{prop}: 
Eq. \ref{2.2} can be solved in the class $S^0(m)$, with $m(x,p)= (1+|p|^2)^{1/2}$.
In particular, the principal symbol for $L$ is given by
$$L_0(x,p)=|p|\tanh(D(x)|p|)$$
and its invariant sub-principal symbol is 0, so that, for Weyl quantization
\begin{equation}
\label{2.16}
L^w(x,hD_x;h)=\Op^w(|p|\tanh(D(x)|p|)+{\cal O}(h^2)
\end{equation}
\end{prop}
 
We notice that $L_0(x,p)\sim |p|$ as $|p|\to\infty$, which expresses that $L=L_{11}:
H^{3/2}(\Gamma^+)\to H^{1/2}(\Gamma^+)$ has a loss of smoothness equal to 1.

We proceed to compute $L_{12}$ in a similar way. Its principal symbol  is given by
$$Q_0(x,p)={1\over\cosh(D(x)|p|)}$$ 
We notice that $Q_0(x,p)$ is smooth in $p$ and decays exponentially as $|p|\to\infty$, 
so in particular $L_{12}:
H^{1/2}(\Gamma^-)\to H^{1/2}(\Gamma^+)$ is bounded. Operator $L_{12}$ turns out also to be smoothing in the classical sense.

Since we have constructed $L(x,p;h)$ we can already solve
$\widehat L\phi_0(x)=\omega^2\phi_0(x)+f^+(x)$, whenever we can invert $\widehat L-E$, $E=\omega^2$. 

\section{Semi-classical Maupertuis-Jacobi correspondence}

Let $H^A(x,p), H^B(x,p)$ be two smooth Hamiltonians possessing a common
regular energy surface $\Sigma=\{H^A=E^A\}=\{H^B=E^B\}$. Then $H^A,H^B$ have same the integral curves 
$(x(t^A),p(t^A))=(x(t^B),p(t^B))$ on $\Sigma$, up to a reparametrization of time. Then we say that 
the pair $(H^A,E^A)$ and $(H^B,E^B)$ satisfy Maupertuis-Jacobi correspondence; we shall
use this property to solve Green function or scattering problem for $L$. 

In \cite{DoRo}, \cite{DoMiRo} we proved the following. Recall 
$L^w(x,hD_x;h)$ from \ref{2.16}, 
with principal symbol 
$L_0(x,p)=|p|\tanh(D(x)|p|)$. Then there is a one parameter family $({\cal H}(E), L_0 )_{E> 0}$ of Hamiltonians satisfying
Maupertuis-Jacobi correspondence at energies $(1, E)$, with ${\cal H}(E)$ is a smooth conformal metric on ${\bf R}^2$
of the form ${\cal H}(x,p;E)=G(x,E)p^2$.
This implies the factorization
\begin{equation}
\label{3.1}
\begin{aligned}
&K_0(x,p;E)=G(x,E)p^2-1=\cr
&C_0(x,p;E)^2(L_0(x,p)-E)\cr
\end{aligned}
\end{equation}
with $C_0(x,p;E)$ elliptic. We carry on this factorization to include higher order terms as follows: 
let $C(x,p;E,h)=C_0(x,p;E)-ihC_1(x,p;E)-h^2C_2(x,p;E)+\cdots$, with real $C_1$
we compute 
\begin{equation}
\label{3.2}
\begin{aligned}
&K^w(x,hD_x;E,h)=C^{w*}(L^w(x,hD_x;h)-E)C^{w}=\cr
&\Op^w\bigl(K_0+h^2K_2+\cdots\bigr)\cr
\end{aligned}
\end{equation}
and find with $C_1=0$
\begin{equation}
\label{3.3}
\begin{aligned}
&K(x,p;E,h)\equiv C_0^2(L_0-E)+h^2\bigl(C_0^2L_2-\widetilde C_0(E)-\cr
&2C_0(L_0-E)\re C_2\bigr)\cr
\end{aligned}
\end{equation}
with $\widetilde C_0(E)$ a smooth function depending on $C_0$ and $L_0-E$ and their derivatives up to order 2.
so that mod ${\cal O}(h^3)$
\begin{equation}
\label{3.4}
K(x,p;E,h)=K_0+h^2\bigl({K_0(E)L_2\over L_0-E}-\widetilde C_0(E)\bigr)
\end{equation}
with $K_0(x,p;E)=G(x,E)p^2-1$ as in \ref{3.1}. So conjugating $L^w(x,hD_x;h)-E$ by an elliptic $h$-PDO
$C^w(x,hD_x;E,h)=\widehat C=\Op^w(C_0(E)-h^2 C_2(E)+\cdots)$
with real $C_2(E)$, we get a Weyl symbol of the form $G(x,E)p^2-1$ mod ${\cal O}(h^2)$, i.e. a Helmoltz type operator.
Now we can rewrite \ref{3.1} as 
$$G(x,E)p^2-1=\bigl(g(x,E)|p|-1\bigr)\bigl(g(x,E)|p|+1\bigr)$$
and complete as above the factorization of the symbol $K(E;h)=K_0(E)+h^2K_2(E)+\cdots$ outside $p=0$, we get
\begin{equation}
\label{3.8}
K^w=D^{w*}\Op^w\bigl(g(x,E)|p|-1+h^2G_2(x,p;E)+..\bigr)D^w
\end{equation}
with smooth symbols outside $p=0$; more precisely with $D_0(x,p;E)=\sqrt{g(x,E)|p|+1}$, and $\widetilde D_0$ computed
as in \ref{3.4} when replacing $C_0$ and $L_0-E$ by $D_0$ and $g(x,p)|p|-1$ resp., we have 
$D(x,p;E,h)=D_0-h^2D_2(x,p)+\cdots$, $\re D_2=0$, and 
$$G_2(x,p,E)=\bigl(g(x,E)|p|+1\bigr)^{-1}\bigl(K_2(x,p,E)+\widetilde D_0\bigr)$$  
Combining \ref{3.4} and \ref{3.8} we have a two-sided factorization of $L-E$ to $\Op^w\bigl(g(x,E)|p|-1+h^2G_2(x,p;E)+\cdots\bigr)$
outside any neighborhood of $p=0$. 

From the conformal metric we can pass again to a Schr\"odinger type operator by Maupertuis-Jacobi correspondence. 
Namely in \cite{DoRo} we proved the following.
Given $L_0$, there is a one parameter family $({\cal K}_E, L_0 )_{E> 0}$ of Hamiltonians satisfying
Maupertuis-Jacobi correspondence at energies $(0, E)$, with ${\cal K}_E$ is Schrodinger or Helmholtz operator
${\cal K}_E(x, p)=p^2-V(x,E)$.
Moreover $V(x;E)={Z^2(ED(x))\over D(x)^2}$ where $Z(s)=z$ is the positive root of 
$s=z\tanh z$. 
This implies as above the factorization
$p^2-V(x,E)=F_0(x,p;E)^2(L_0(x,p)-E)$
with $F(x,p;E)$ elliptic, and as before
this factorization extends to the full Weyl symbol as 
\begin{equation}
\label{3.11}
\begin{aligned}
&{\cal K}=p^2-V(x,E)+h^2V_2(x,p,E)+\cdots=\cr
&F^*\sharp(L_0(x,p)-E+h^2 L_2(x,p)+\cdots)\sharp F\cr
\end{aligned}
\end{equation}
where $\sharp$ denote composition of Weyl symbols.

\section{Green functions and Scattering theory}

Let $H$ be a self-adjoint $h$-PDO with principal symbol $H_0$ such that $H_0^{-1}(E)$ is a regular energy surface.
The forward parametrix of $H$ is defined resp. by
$E_f={i\over h}\int_0^\infty e^{-it(H-E)/h}\,dt$
It solves (formally) $(H-E)E_f=E_f(H-E)=\id$.
The kernel of such operators make sense as oscillating integrals, and can be also defined by boundary value of the
resolvent $R(E\mp i0)$. We give realizations of these operators acting on some Lagrangian distributions as 
Maslov canonical operators of a precise type.\\

\noindent {\it a) Asymptotic Green function with limiting absorption condition}\\

We present next our formula for the forward parametrix $E_f={i\over h}\int_0^\infty e^{-it(H-E)/h}\,dt$. 
Assume that Cauchy data is a Lagrangian distribution given by Maslov canonical operator
\begin{equation}
\label{4.3}
f(x,h)=\bigl[K^h_{(\Lambda, d\mu)}A\bigr](x;h)
\end{equation}
acting on a smooth density $A$. In particular if
$\Lambda=\{(x,p)\in{\bf R}^{2n}: x=x_0, p\in{\bf R}^n\}$
is the conormal bundle to the point $x_0$ as above, then $f(x;h)$ is a localized (in space) semi-classical distribution of the form
\begin{equation*}
\begin{aligned}
&f(x;h)=(2\pi h)^{-n/2}V({x-x_0\over h})\cr 
&V(y)=e^{i\pi n/4}\int_{{\bf R}^n}e^{ipy}A(p)\,dp\cr
\end{aligned}
\end{equation*}
where $A\in C^\infty_0(\Lambda)$. 

\begin{definition} Given $f\in L^2$ as above,
a semi-classical Lagrangian distribution $u(x;h)$ is called an ``asymptotic Green function with limiting absorption condition'' for 
$\widehat H u=f$ if there is a family $u_{\e} (x;h)$ such that 
$$(\widehat H-i\e )u_{\e }=f+{\cal O}(h^\infty)$$ 
uniformly in $\e $, and $\|u-u_{\e} \|\to 0$ in all local (semi-classical) Sobolev norm,
locally uniformly in $x$ in $h$ as $\e \to0$. 
\end{definition}
This condition is the semiclassical analogue of well-known Limiting Absorption Principle for
Helmholtz operator.
Assume on the other hand that the characteristic variety $H(x,p)=0$ intersects tranversally $\Lambda$
along an isotropic manifold $\ell_0$.
If $\Lambda_+=\bigcup_{t\geq0}\Lambda_t$, $\Lambda_t=g_H^t(\ell_0)$ 
being the Hamiltonian flow out of $\ell_0$, then the pair $(\Lambda,\Lambda_+)$ define
a Lagrangian intersection in the sense of \cite{MelUh}. 

Choose an initial point $\alpha^*\in \ell_0$, and its image $\alpha_t^*=g_H^t(\alpha^*)$ on $\Lambda_t$ by the Hamiltonian flow.
Next define a measure on $\Lambda_t$ as
$d\mu_t=(g_H^{t})^{*-1}d\mu$, where $d\mu$ is a measure on
$\Lambda$, and a measure $d\mu_+$ on $\Lambda_+$. 
Let $\widetilde A$ be te solution of the transport equation 
along the trajectory $g_H^t(m_0)=(x(m_0,t),p(m_0,t))$
issued from $m_0=(x_0,p_0)\in\Lambda$ with initial condition $A|_{t=0}=A(x_0,p_0)$
where we recall $A\in C_0^\infty(\Lambda)$ from \ref{4.3}. 
Let us denote the restrictions of the function $\widetilde A$
to Lagrangian manifolds $\Lambda_+$ and $\Lambda_t$ by $A_+$
and $A(t)$ resp. Let $\rho\in C^\infty(\Lambda)$ and $\theta\in
C_0^\infty({\bf R})$ be suitable cutoff functions,
$\theta$ being interpreted as a function on
$\Lambda_+$, if we think of $t$ as the time variable along a trajectory of Hamilton vector field $X_H$, 
and $\rho$ as a function on $\Lambda_t$, constant along the trajectories of $X_H$.

With these notations we recall from \cite{AnDoNaRo} the following structure result:

\begin{theo}
Assume $H$ has real principal symbol, the energy surface $H=0$ is regular,
$X_H$ is transverse to $\Lambda$ in
$\ell_0\subset\{H=0\}$ and $\ell_0$ is non-trapping with respect to position variable.
Then the leading term of $E_f$ is of the form (letting $\e \to 0$):
\begin{equation}
\label{4.14}
\begin{aligned}
u&=K^h_{(\Lambda,d\mu)}\bigl[{(1-\rho)A\over H}\bigr]+\cr
&\bigl({2i\pi\over h}\bigr)^{1/2}
K^h_{(\Lambda_+,d\mu_+)}\bigl[(1-\theta(t_0))A(t_0)\bigr]\cr
&+{i\over h}\int_0^{t_0}\exp\bigl[i\int_{\alpha^*}^{g_H^t(\alpha^*)} p\,dx\bigr]\cr
&K^h_{(\Lambda_t,d\mu_t)}\rho\theta(t) A(t)\,dt+{\cal O}(h^{1/2})\cr
\end{aligned}
\end{equation}
where the integral in the exponent is calculated along the trajectory of Hamilton vector field $X_H$.

\end{theo}
See also Van Vleck formula \cite{CdV} (Sect.9,Theorem 59), and \cite{M}, \cite{Ku}. 

We apply this Theorem to $L-E$, or rather its normal form ${\cal H}=g(x,E)|p|-1+h^2G_2(x,p;E)$ given by \ref{3.8}.
If the localized source $f$ is of the form \ref{4.3} and $\widehat C=C^w(x,hD_x;E,h)$ is an elliptic $h$-PDO, 
so does $C^w(x,hD_x;E,h)f$, because of the commutation formula 
\begin{equation}
\label{4.16}
\widehat C K^h_{(\Lambda,d\mu)}A=K^h_{(\Lambda,d\mu)}\bigl(C_0(x,p;E)|_{\Lambda} A\bigr)(1+{\cal O}(h))
\end{equation}
The advantage of using ${\cal H}$ is that $g(x,E)|p|$ (reversible Finsler matric) is positively homogeneous 
of degree 1, and the corresponding Hamiltonian system has very nice properties. 
Using (4.16), it is clear that all hypotheses of Theorem 4.2 hold, so we have the asymptotic solution of 
$(\widehat L-E)u=f$ mod ${\cal O}(h)$. Note that one could also work directly with $L-E$ \cite{DoNa}.\\

\noindent {\it b) Scattering theory}\\

In this Section we consider semi-classical
Scattering Theory and prove limiting absorption principle as in \cite{RoTa}, \cite{GeMa}. 
To this end, we will convert again the Hamiltonian $L(x,hD_x;h)$ to Schr\"odinger type Hamiltonian ${\cal K}(x,hD_x;E,h)$
with symbol \ref{3.11}. 

As $D(x)\to D_0$ as $|x|\to\infty$, we have $V(x,E)\to{Z^2(ED_0)\over D_0^2}=v(E)$ as $|x|\to\infty$, 
so we rewrite \ref{3.11} as 
$p^2-W(x,E)-v(E)+h^2V_2(x,p;E)$
with $W(x,E)\to0$ as $|x|\to\infty$. If 
\begin{equation*}
|\partial_x^\alpha (D(x)-D_0)|\leq C_\alpha\langle x\rangle^{-|\alpha|-\rho}
\end{equation*}
where $\langle x\rangle=(1+x^2)^{1/2}$ and $\rho>0$ (i.e. $x\mapsto D(x)-D_0$ is long range)
we have also
\begin{equation}
\label{4.19}
|\partial_x^\alpha W(x,E)|\leq C'_\alpha\langle x\rangle^{-|\alpha|-\rho}
\end{equation} 
locally uniformly in $E\in I$.

Assume also that the energy surface $\Sigma_{E_0}=\{L_0(x,p)=E_0\}$, $E_0\in I$, is non trapping, i.e.
\begin{equation}
\label{4.20}
\forall (x,p)\in \Sigma_E, \ g^t_{L_0}(x,p)\to\infty \ \hbox{as} \ t\to+\infty \ \hbox{or} \ t\to-\infty
\end{equation}

Using Mourre estimates as in \cite{RoTa}, \cite{GeMa} we get the following limiting absorption principle

\begin{theo} 
Let $R(E\pm i0,h)$ be the boundary values of the resolvent 
$R(z,h)=(L(x,hD_x;h)-z)^{-1}$
for $|E-E_0|$ small enough. Assume \ref{4.19}, \ref{4.20}. 
If $s>1/2$, then  
$$\|\langle x\rangle^{-s}R(E\pm i0,h)\langle x\rangle^{-s}\|={\cal O}(h)^{-1}$$
\end{theo}

The first step is to prove the existence of a $h$-PDO ${G}^w$ conjugate to ${\cal K}^w$, i.e.
that satisfies Mourre estimate
\begin{equation}
\label{4.21}
\chi({\cal K}^w)\frac{i}{h}\big[\widehat{\cal K}^w,{G}^w\bigr]\chi({\cal K}^w)\geq h\chi^2({\cal K}^w)
\end{equation}
uniformly for $h>0$ small enough. Here $\chi$ is a smooth cut-off function equal to 1 on $I$, and ${\cal K}$
is as \ref{3.11}. 
This follows from the non-trapping conditions \ref{4.19}, \ref{4.20} as in \cite{GeMa}. It remains to prove such an estimate
for $L-E=F^{w-1*}{\cal K}^wF^{w-1}$ in \ref{3.11}. First we observe that 
at the level of the principal symbols, 
$$\{L_0-E,{G}\}=F_0^{-2}\{{\cal K}_0,{G}\}+\{F_0^{-2},{G}\}{\cal K}_0$$
By \ref{4.19}, \ref{4.20}, the first term is bounded below from 0 for $E\in I$, while 
$\{F_0^{-2},{G}\}={\cal O}(1)$ uniformly in $E\in I$, and ${\cal K}_0\to 0$ as $|I|\to 0$.
So $\{L_0,{G}\}$ is bounded from below from 0 if $|I|$ is small enough. At the level of operator
$\frac{i}{h}\big[{\cal K}^w,{G}^w\bigr]$, we modify the Hilbertian norm on $L^2$ microlocally
near $\{L=E\}$ by an elliptic weight so that Mourre estimate \ref{4.21} holds in the new Hilbert space.
Then we end up the proof as in \cite{GeMa}.

\end {document}